\documentclass{article}
\usepackage{hq2kpro,epsfig}
\begin{document}
\title{ THE LIGHT SCALARS, ESPECIALLY THE SIGMA, IN CHARM DECAY AND ELSEWHERE}
\author{N.A. T\"ornqvist and A. D. Polosa
\\ {\em Physics Department, POB 9, FIN-00014, University of
Helsinki, Finland}
\\ } \maketitle \baselineskip=11.6pt
\begin{abstract}
We discuss how the lightest scalars, in particular the broad
$\sigma$ resonance, can be understood as unitarized $q\bar q$
states within a unitarized quark model (UQM). The bare $q\bar q$
scalars are  strongly distorted  by hadronic mass shifts, and the
$u\bar u+d\bar d$ state becomes a very broad resonance, with its
pole at 470-i250 MeV. This is the sigma meson required by models
for spontaneous breaking of chiral symmetry. We also discuss the
less well known phenomenon  that with a large coupling there can
appear  two physical resonance poles on the second sheet although
only one bare quark-antiquark state is put in. The $f_0(980)$ and
$f_0(1370)$ resonance poles can thus be two manifestations of the
same $s\bar s$ quark  state. Both of these are dominant in the
E791 Dalitz plot of $D_s\to 3\pi$, where $s\bar s$ intermediate
states should be dominant. Recently this light $\sigma$ has
clearly been observed in $D\to\sigma\pi\to3\pi$ by the E791
experiment at Fermilab. We discuss how this decay channel can be
predicted in a Constituent Quark Meson Model (CQM), which
incorporates heavy quark and chiral symmetries.
\end{abstract}
%
\baselineskip=14pt

\def \gam {\frac{ N_f N_cg^2_{\pi q\bar q}}{8\pi} }
\def \gam {\frac{ N_f N_cg^2_{\pi q\bar q}}{8\pi} }
\def \gamm {N_f N_cg^2_{\pi q\bar q}/(8\pi) }
\def \be {\begin{equation}}
\def \ba {\begin{eqnarray}}
\def \ee {\end{equation}}
\def \ea {\end{eqnarray}}
\def \gap {{\rm gap}}
\def \gapp {{\rm \overline{gap}}}
\def \gappp {{\rm \overline{\overline{gap}}}}
\def \im {{\rm Im}}
\def \re {{\rm Re}}
\def \Tr {{\rm Tr}}
\def \P {$0^{-+}$}
\def \S {$0^{++}$}
\def \uu {$u\bar u$}
\def \dd {$d\bar d$}
\def \ss {$s\bar s$}
\def \qq {$q\bar q$}
\def \qqq {$qqq$}
\def \lsm {L$\sigma$M}
\def \sig {$\sigma$}
\def \gam {\frac{ N_f N_cg^2_{\pi q\bar q}}{8\pi} }
\def \gamm {N_f N_cg^2_{\pi q\bar q}/(8\pi) }
\def \be {\begin{equation}}
\def \ba {\begin{eqnarray}}
\def \ee {\end{equation}}
\def \ea {\end{eqnarray}}
\def\bea{\begin{eqnarray}}
\def\eea{\end{eqnarray}}
\def \gap {{\rm gap}}
\def \gapp {{\rm \overline{gap}}}
\def \gappp {{\rm \overline{\overline{gap}}}}
\def \im {{\rm Im}}
\def \re {{\rm Re}}
\def \Tr {{\rm Tr}}
\def \P {$0^{-+}$}
\def \S {$0^{++}$}
\def\zpp{$0^{++}$}
\def\fz{$f_0(980)$}
\def\az{$a_0(980)$}
\def\Kz{$K_0^*(1430)$}
\def\fzz{$f_0(1300)$}
\def\fzzz{$f_0(1200-1300)$}
\def\azz{$a_0(1450)$}
\def\ss{$ s\bar s $}
\def\uu{$u\bar u+d\bar d$}
\def\qq{$q\bar q$}
\def\KK{$K\bar K$}
\def\sig{$\sigma$}
\def\lsim{\;\raise0.3ex\hbox{$<$\kern-0.75em\raise-1.1ex\hbox{$\sim$}}\;}
\def\gsim{\raise0.3ex\hbox{$>$\kern-0.75em\raise-1.1ex\hbox{$\sim$}}}

\section{Introduction}
This talk is mainly based on earlier papers~\cite{NAT1,NAT2} on
the light scalars and on a more recent one \cite{gatto} on the
\sig\ in charm decay, including a few new comments. First we shall
discuss the evidence for the light \sig\ and explain how one can
understand the controversial light scalar mesons with unitarized
quark  model (UQM), which includes most well established
theoretical constraints:
\begin{itemize}
\item Adler zeroes as required by chiral symmetry,
\item all light two-pseudoscalar (PP) thresholds with flavor symmetric couplings in a coupled channel framework
\item physically acceptable analyticity, and
\item unitarity.
\end{itemize}  A unique
feature of this model is that it simultaneously describes the  whole scalar
nonet and one obtains a good representation of a large
set of relevant data. Only six parameters,
which all have a  clear physical interpretation, are
needed, such as an overall coupling constant, the bare mass
of the $u\bar u$ or $d\bar d$ state,
the extra mass for a strange quark ($m_s-m_u=100$ MeV),
a cutoff parameter ($k_0=0.56$ GeV/c).

After describing our understanding of the $q\bar q$ nonet, we
discuss the recently measured $D\to\sigma\pi\to 3\pi$ decay, where
the \sig\ is clearly seen as the dominant peak.

\section{The problematic scalars and the existence of the \sig }

The interpretation of the nature of lightest scalar mesons has
been controversial for long. There is no general agreement on
where are the  $q\bar q$ states, is there a glueball among the
light scalars, are some of the scalars multiquark or $K\bar K$
bound states? As for the $\sigma$, authors do not even agree on
its existence as a fundamental hadron, although the number of
supporters is growing rapidly.
\begin{figure}[t!]
\begin{center}
\epsfxsize=12 cm \epsfysize=6. cm \epsffile{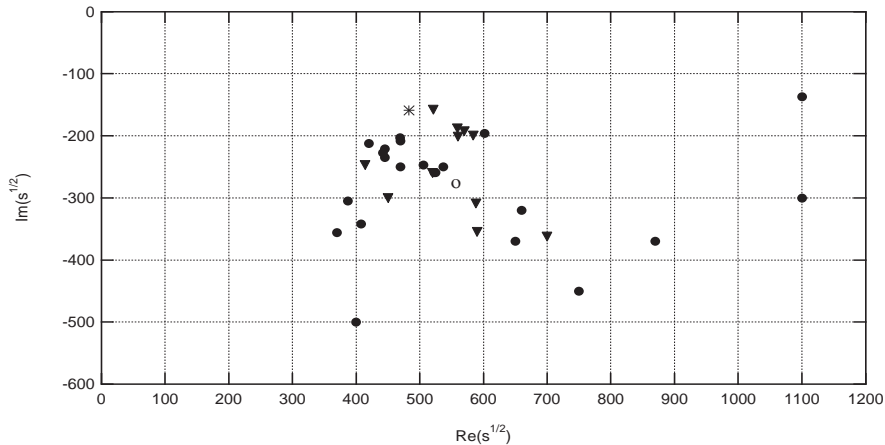}
\caption{\label{fig:poles} \it The pole positions of the \sig\
resonance, as listed by the PDG \cite{pdg2000} under
$f_0(400-1200)$ or \sig\ (filled circles), plotted in the complex
energy plane (in units of MeV). The triangles represent the mass
and width parameters (plotted as $m-i\Gamma /2$), which were
reported at this meeting. We could not here distinguish between
pole and Breit-Wigner parameters. The star is the $m-i\Gamma /2$
point obtained from the recent E791 experiment \cite{E791} on
$D\to\sigma \pi\to 3\pi$ ($m_\sigma=478$ MeV, $\Gamma_\sigma =
324$ MeV) while the open circle is that obtained by the CLEO
analysis of $\tau\to\sigma\pi\nu\to 3\pi\nu$ \cite{CLEO}.}
\end{center}
\end{figure}
A light scalar-isoscalar meson (the \sig ), with a mass of twice
the constituent  $u,d$ quark mass, or $\approx 600$ MeV, coupling
strongly to $\pi\pi$ is of importance in all
Nambu--Jona-Lasinio-like (NJL-like) models for dynamical breaking
of chiral symmetry. In these models the $\sigma$ field obtains a
vacuum expectation value, i.e., one has a \sig\  $q\bar q$
condensate in the vacuum, which is crucial for the understanding
of all hadron masses, as it explains in a simple way the
difference between the light constituent and chiral quark mass.
 Then most of the nucleon mass is generated by its coupling to
the $\sigma$, which acts like an effective Higgs-like
 boson for the hadron spectrum.

In Fig. 1 we have plotted with filled circles the results of 22
different analyses on the \sig\ pole position, which are included
in the  2000 edition of the Review of Particle Physics
\cite{pdg2000} under the entry $f_0(400-1200)$ or \sig.  Most of
these find a \sig\ pole position near 500-i250 MeV.

Also, at  a recent meeting in Kyoto \cite{kyoto} devoted to the
$\sigma$, many groups reported preliminary analyzes, which find
the \sig\ resonance parameters in the same region. These are
plotted as triangles in Fig. 1. Here it was not possible to
distinguish between Breit-Wigner parameters and pole positions,
which of course can differ by several 100 MeV for the same data.
 It must also be noted that many of the triangles in Fig. 1 rely on the same
raw data and come from preliminary analyzes not yet published.

We also included in Fig. 1 (with a star) the \sig\ parameters
obtained from the recent E791 Experiment at Fermilab \cite{E791},
where 46\% of the $D^+\to3\pi$ Dalitz plot is $\sigma\pi$.  The
open circle in the same figure represents the \sig\ parameters
extracted from the CLEO analysis of $\tau\to\sigma\pi\nu\to
3\pi\nu$ \cite{CLEO}.
\section{The NJL and the linear sigma model}

The NJL model is an effective theory which is believed to be
related to QCD at low energies, when one has integrated out the
gluon fields. It involves a linear realization of chiral symmetry.
After bosonization of the NJL model one finds essentially the
linear sigma model (\lsm ) as an approximate effective theory for
the scalar and pseudoscalar meson sector.

About  30 years ago Schechter and Ueda \cite{u3u3}  wrote down the
$U3\times U3$ \lsm\ for the meson sector involving a scalar and a
pseudoscalar nonet. This (renormalizable) theory has only 6
parameters, out of which 5 can be fixed by the pseudoscalar masses
and decay constants ($m_\pi,\ m_K, \ m_{\eta^\prime},\ f_\pi, \
f_K$). The sixth parameter for the OZI rule violating 4-point
coupling must be small. One can then predict, with no free
parameters, the tree level scalar masses \cite{lsm}, which turn
out to be not far from the lightest experimental masses, although
the two quantities are \underline{not} exactly the same thing but
can differ for the same model and data by over 100 MeV.

The important thing is that the scalar masses are predicted to be
near the lightest experimentally seen scalar masses, and not in
the 1500 MeV region where many authors want to put the lightest
$q\bar q$ scalars. The \sig\ is predicted \cite{lsm} at 620 MeV
with  a very large width ($\approx 600$ MeV), which well agrees
with Fig. 1. The $a_0(980)$ is predicted at 1128 MeV,  the
$f_0(980)$ at 1190 MeV, and the $K^*_0(1430)$ at 1120 MeV, which
is surprisingly good considering that loop  effects are large.

\section{ Understanding the S-waves within a unitarized quark model (UQM)}
In Figs.~2-4 we show the obtained fits to the $K\pi$, $\pi\pi$ S
-waves and to the  \az\ resonance peak in $\pi\eta$. The Partial
Wave Amplitude (PWA) in the case of one \qq\ resonance, such as
the $a_0(980)$, can be written as:
\begin{equation} A(s)=-
\frac{Im\Pi_{\pi\eta}(s)}{[m_0^2+Re\Pi (s)-s +iIm\Pi (s)]},
\end{equation}
where: \bea \label{PWA} Im\Pi (s)&=&\sum_i Im\Pi_i(s) \label{impi}
=-\sum_i \gamma_i^2(s-s_{A,i})\frac{k_i}{\sqrt s}e^{-k_i^2/k_0^2}
           \theta(s-s_{th,i})\ ,\nonumber \\
Re\Pi (s)&=&\frac 1\pi{\rm P.V.}\int^\infty_{s_{th,1}} \frac{Im\Pi
(s)}{s'-s} ds' \ . \nonumber \eea Here the coupling constants
$\gamma_i$ are related by flavour symmetry and OZI rule, such that
there is only one over all parameter $\gamma$. The $s_{A,i}$ are
the positions of the Adler zeroes, which  are  near $s=0$. Eq. (1)
can be looked upon as a more general Breit-Wigner form, where the
mass parameter is replaced by an $s$-dependent function, ``the
running mass" $m_0^2+ Re\Pi (s)$.

In the flavourless channels the situation is a little more
complicated than in Eq. (1) since one has both \uu\ and \ss\
states, requiring a two dimensional mass matrix (see
Ref.~\cite{NAT2}). Note that the sum runs over all light PP
thresholds, which means three for the \az : $\pi\eta ,\ K\bar
K,\pi\eta'$ and three for the \Kz : $ K\pi ,\ K\eta ,\ K\eta' $,
while for the $f_0$'s there are five channels: $\pi\pi ,K\bar K ,\
\eta\eta ,\ \eta\eta' ,\ \eta'\eta'$.

In Fig.~5 we show, as an example, the running mass,
$m_0^2+Re\Pi(s)$, and the width-like function, $-Im\Pi(s)$, for
the I=1 channel. The crossing point of the running mass with $s$
gives the $90^\circ$  mass of the \az.
 The magnitude of the \KK\ component in the \az\ is determined by
$-\frac d{ds}Re\Pi(s)$, which is large in the resonance region
just below the \KK\ threshold. These functions fix the PWA of Eq.
(1) and Fig.~3. In Fig. 6 the running mass and width-like function
for the strange channel are shown. These fix the shape of the
$K\pi$ phase shift and absorption parameters in Fig. 1. As can be
seen from Figs.~1-3, the model gives a good description of the
relevant data.

In Ref.~\cite{NAT2} the \sig\ was missed because only poles
nearest to  the physical region were looked for, and the
possibility of the resonance doubling phenomenon, discussed below,
was overlooked. Only a little later we realized with Roos
\cite{NAT1} that two resonances (\fz\ and $ f_0(1370)$) can emerge
although only one $s\bar s$ bare state is put in. Then we had to
look deeper into the second sheet and found the broad \sig\ as the
dominant singularity at low mass.

In fact, it was pointed out by Morgan and Pennington~\cite{morgan}
that for each \qq\ state there are, in general, apart from the
nearest pole, also  image poles, usually located far from the
physical region. As explained in more detail in Ref.~\cite{NAT1},
some of these can (for a large enough coupling and sufficiently
heavy threshold) come so close to the physical region that they
make new resonances. And, in fact, there are more than four
physical poles  with different isospin, in the output spectrum of
the UQM model, although only four bare states, of {\it the same
nonet}~\cite{NAT1},  are put in!. The \fz\ and the \fzz\ of the
model thus turn out to be  two manifestations of the same \ss\
state. (See Ref. \cite{NAT1} for details). There can be two
crossings (see Fig. 7) with the running mass $m_0^2+ Re\Pi(s)$,
one near the threshold and another at higher mass, and each one is
related to a different pole at the second sheet (or, if the
coupling is strong enough, the lower one could even become a bound
state pole, below the threshold, on the first sheet).

 Similarly the \az\ and the  \azz\ could be two manifestations of the $u\bar d$ state.
Only after  realizing that this resonance doubling is
important the light and broad \sig\ was found in the model
\cite{NAT1}.
In Table 1 we list the pole positions  of the six relevant poles,
all manifestations of the same $q\bar q$ nonet.

\begin{table}[t] \caption{
The pole positions of the
resonances in the S-wave  $ PP \to PP $ amplitudes~$^{1)}$. 
The first resonance is the \sig\ which we name here $f_0(\approx
500)$. The two following are both different manifestations of the
same $s\bar s$ state. The last entry is similarly an image pole of
the \az , which in an improved fit could represent the
$a_0(1450)$.
 The mixing angle $\delta_S $ for the
$f_0(\approx 500)$ or $\sigma$  is with respect to
$u\bar u +d\bar d$, while for the two heavier
$f_0$'s it is with respect to $s\bar s$. }
\vspace{ 0.5cm}
\begin{center}
\begin{tabular}{|c|c|c|c|}
\hline
resonance&$s_{\rm pole}^{1/2}$&$\delta_{S,pole}$&Sheet\\
\hline
$f_0(\approx 500)$&$470-i250           $&$(-3.4+i1.5)^\circ $&II    \\
$f_0(980)        $&$1006-i17          $&$(0.4+i39)^\circ   $&II    \\
$f_0(1370)       $&$1214-i168         $&$(-36+i2)^\circ    $&III,V \\
$K_0^*(1430)     $&$1450-i160         $&  -                 &II,III\\
$a_0(980)        $&$1094-i145         $&  -                 &II    \\
$a_0(1450)?      $&$1592-i284         $&  -                 &III   \\
\hline
\end{tabular}
\vspace{0.5cm}\end{center}
\end{table}

Another important effect that the model can explain is the large
mass difference between the $a_0$ and $K^*_0$. Because of this
large mass splitting many authors argue that the $a_0(980)$ and
$f_0(980)$ are not \qq\ states, since in addition to being very
close to the \KK\ threshold, they are much lighter than the first
strange scalar, the  $K^*_0(1430)$. Naively one expects a mass
difference between the strange and nonstrange meson to be of the
order of the strange-nonstrange quark mass difference, or a little
over 100 MeV.

Figs. 5 and 6 explain why  one can easily understand this large
$K^*_0(1430)-a_0(980)$ mass splitting as a secondary effect of the
large pseudoscalar mass splittings, and because of the large mass
shifts coming from the loop diagrams involving the PP thresholds.
If one puts Figs. 4 and 5 on top of each other one  sees that  the
3 thresholds $\pi\eta,\ K\bar K,\ \pi\eta$ all lie relatively
close to the $a_0(980)$, and all 3 contribute to a large mass
shift. On the other hand, for the $K^*_0(1430)$, the $SU3_f$
related thresholds ($K\pi,\ K\eta'$) lie far apart from the
$K^*_0$, while the $K\eta$ nearly decouples because of the
physical value of the pseudoscalar mixing angle.

This large mass of the $K^*_0(1430)$ is also one of the reasons
why some authors want to have a lighter strange meson, the
$\kappa$, near 800 MeV. Cherry and Pennington \cite{cherry}
recently have strongly argued against its existence. But, we heard
Carla G\"obel in her talk describing some evidence for such a
light $\kappa$ in the E791 data for $D^+\to K^-\pi^+\pi^+$. Here
the signal is much less evident, since, differently from the case
of the $\sigma$, which is seen as a clear peak over background,
the inclusion of a $\kappa$ only improves the $\chi^2$ in the
region dominated by the $K^*(890)$. Perhaps one should  try a more
sophisticated Breit-Wigner amplitude for the S-wave, as that in
Eq. (1), before one can make more definite statements about the
$\kappa$. Possibly such a light $\kappa$ could be understood in
connection to the resonance doubling phenomenon discussed above
and which we discussed with Roos \cite{NAT1}.

\section{$D\to\sigma\pi\to 3\pi$}

The recent experiments studying charm decay to light hadrons are
opening up a new experimental window for understanding light meson
spectroscopy and especially the controversial scalar mesons, which
are copiously produced in these decays.

In particular we refer to the E791 study of the $D\to 3\pi$ decay
\cite{E791} where it is shown how adding an intermediate scalar
resonance with floating mass and width in the Monte Carlo program
simulating the Dalitz plot densities, allows for an excellent fit
to data provided the mass and the width of this scalar resonance
are $m_\sigma\simeq 478$ MeV and $\Gamma_\sigma\simeq 324$ MeV.
This resonance is a very good candidate for the $\sigma$. To check
this hypothesis we adopt the E791 experimental values for its mass
and width and using a Constituent Quark Meson Model (CQM) for
heavy-light meson decays \cite{rass} we compute the
$D\to\sigma\pi$ non-leptonic process via {\it factorization}
\cite{WSB}, taking the coupling of the $\sigma$ to the light
quarks from the Linear sigma Model \cite{volkoff}. In such a way
one is directly assuming that the scalar state needed in the E791
analysis could be the quantum of the $\sigma$ field of the Linear
sigma Model. According to the CQM model and to factorization, the
amplitude describing the $D\to\sigma\pi$ decay can be written as a
product of the semileptonic amplitude $\langle
\sigma|A^\mu_{(\bar{d}c)}(q)|D^+\rangle$, where $A^\mu$ is the
axial quark current, and $\langle \pi|A_{\mu(\bar{u}d)}(q)|{\rm
VAC}\rangle$. The former is parameterized by two form factors,
$F_1(q^2)$ and $F_0(q^2)$, connected by the condition
$F_1(0)=F_0(0)$, while the latter is governed by the pion decay
constant $f_\pi$. As far as the product of the two above mentioned
amplitudes is concerned, only the form factor $F_0(q^2)$ comes
into the expression of the $D\to\sigma\pi$ amplitude. Moreover we
need to estimate it at $q^2\simeq m_\pi^2$, that is the physically
realized kinematical situation. The CQM offers the possibility to
compute this form factor through two quark-meson 1-loop diagrams
that we call the {\it direct} and the {\it polar} contributions to
$F_0(q^2)$. These quark-meson loops are possible since in the CQM
one has effective vertices (heavy quark)-(heavy meson)-(light
quark) that allow us to {\it compute} spectator-like diagrams in
which the external lines represent incoming or outgoing heavy
mesons while the internal lines are the constituent light quark
and heavy quark propagators.

In Figs. 8 and 9 we show respectively the {\it direct} and the
{\it polar} diagrams for the semileptonic amplitude $D\to\sigma$,
the former being characterized by the axial current directly
attached to the constituent quark loop, the latter involving an
intermediate $D(1^+)$ or $D(0^-)$ state. These two diagrams are
computed with an analogous technique and one finally obtains a
determination of the direct and polar form factors $F_0^{\rm dir,
pol}(q^2)$. The extrapolation to $q^2\simeq m_\pi^2\simeq 0$ is
safe for  the direct form factor while is not perfectly under
control for the polar form factor since the latter is more
reliable at the pole $q^2\simeq m_P^2$, $m_P$ being the mass of
the intermediate state in Fig. 9. We take into account the
uncertainty introduced by this extrapolation procedure and
signaled by the fact that we find $F_0^{\rm pol}(0)\neq F_1^{\rm
pol}(0)$ (computing $F_0$ from the polar diagram with $0^-$
intermediate polar state and $F_1$ from that with intermediate
$1^+$ state). Our estimate for $F_0(0)=F_0^{\rm pol}(0)+F_0^{\rm
dir}(0)=0.59\pm 0.09$ is in reasonable agreement with an estimate
of $F_0(m_\pi^2)=0.79\pm 0.15$ carried out in \cite{dib} using the
E791 data analysis and a Breit-Wigner like approximation for the
$\sigma$.

The meson-quark loops in Figs. 8 and 9 are computed substituting
the meson vertices with the heavy meson field expressions found by
Heavy Quark Effective Theory (HQET) (since CQM incorporates heavy
quark and chiral symmetries) and the quark lines with the
propagators of the heavy and light constituent quarks. The light
constituent mass $m$ is fixed by a NJL-type gap equation that
depends on $m$, and on two cutoffs $\Lambda$ and $\mu$ in a proper
time regularization scheme for the diverging integrals. The
ultraviolet cutoff $\Lambda$ is fixed by the scale of chiral
symmetry breaking, $\Lambda_\chi\simeq 4\pi f_\pi$, and we
consider $\Lambda=1.25$ GeV. The remaining dependence of $m$ on
the choice of the infrared cutoff $\mu$ has an expression similar
to that of a ferromagnetic order parameter, $m(\mu)$ being
different from zero for $\mu$ values smaller than a particular
$\mu_c$, and zero for higher values. When $\mu$ ranges from $0$ to
$300$ MeV, the value of $m$ is almost constant, $m=300$ MeV,
dropping for higher $\mu$ values. A reasonable light constituent
quark mass is certainly $300$ MeV and this clearly leaves a $300$
MeV open window for choosing the infrared cutoff. Enforcing the
kinematical condition for the meson to decay to its free
constituent quarks, which must be possible since the CQM model
does not incorporate confinement, requires $\mu\simeq m$
\cite{rass}. Therefore we pick up the $\mu=300$ MeV value. The
results are quite stable against $10-15\%$ variations of the UV
and IR cutoffs.

The CQM semileptonic $D\to\sigma$ transition amplitude is
represented by the loop integrals associated to the direct and to
the polar contributions. The result of the integral computations
must then be compared with the expression for the hadronic
transition element  $\langle \sigma |A|D\rangle$ and this allows
to extract the desired $F_{0,1}$ form factors. An estimate of the
weight of $1/m_c$ corrections can also be taken into account
\cite{gatto}.

This computation indicates that the scalar resonance described in
the E791 paper can be consistently understood as the $\sigma$ of
the Linear sigma Model. Of course a calculation such as the  one
here described calls for alternative calculations and/or
explanations of the E791 data for a valuable and useful comparison
of point of views on the $\sigma$ nature.

\section{Concluding remarks}
An often raised question is: Why are the mass shifts required by unitarity
so much more important for the scalars than, say, for the vector  mesons?
The answer is very simple, and there are three main reasons:
\begin{itemize}
\item
The scalar coupling to two pseudoscalars is very much larger than
the corresponding coupling for the vectors, both experimentally
and theoretically (e.g., spin counting gives  3 for the ratio of
the two squared couplings).
\item For the scalars the thresholds are S-waves,
giving nonlinear square root cusps in
the $\Pi(s)$ function, whereas for the vectors the thresholds are
P-waves, giving a smooth $k^3$ angular momentum and phase space
factor.
\item
Chiral symmetry constraints, in particular Adler zeroes, are
important for the scalars when analyzing pseudoscalar scattering,
and make, e.g., $\pi\pi\to \pi\pi$ very weak near the thresholds.
In the case when a light scalar is produced in charm decay, as in
E791, these zeroes are less important.
\end{itemize}
\vskip .2cm

One could argue that the two states \fz\ and \az\ are a kind of
$K\bar K$ bound states (see Ref.~\cite{wein}), since these have a
large component of $K\bar K$ in their wave functions. However, the
dynamics of these states is quite different from that of normal
two-hadron bound states. If one wants to consider them as \KK\
bound states, it is the $K\bar K \to s\bar s \to K\bar K$
interaction which  creates their binding energy, not the hyperfine
interaction as in Ref.~\cite{wein}. Thus, although they may spend
most of their time as $K\bar K$, they owe their existence to the
\ss\ state. Therefore, it is more natural to consider the \fz\ and
\fzz\ as two manifestations of the same \ss\ state.

The wave function of the $a_0(980)$ (and $f_0(980)$) can be
pictured as a relatively small core of \qq\ of typical \qq\ meson
size (0.6 fm), which is  surrounded by a much larger standing
S-wave of virtual \KK (see Fig. 10) due to the fact that these
resonances are just below the $K \bar K $ threshold and they
strongly couple to $K\bar K$. This picture also gives a physical
explanation of the narrow width: in order to decay to $\pi\eta$,
the \KK\ component must first virtually annihilate near the origin
to \qq. Then the \qq\ can decay to $\pi\eta$ as an OZI allowed
decay.

Finally, in Sec. 5, we showed that the recent E791 data on
$D\to\sigma\pi$ can be understood in the CQM model assuming the
$\sigma$ to be predominantly a $(u\bar u + d \bar d)/\sqrt{2}$
state, similar to the $\sigma$ of the Linear sigma Model.

\section{Acknowledgements}
NAT and ADP acknowledge support  from EU-TMR programme, contract
CT98-0169. The authors are also grateful to A. Deandrea, R. Gatto
and G. Nardulli for useful discussions.

{\twocolumn[\hsize\textwidth\columnwidth\hsize\csname
@twocolumnfalse\endcsname ]
\begin{figure}
\vspace{4cm}
\begin{center}
\includegraphics{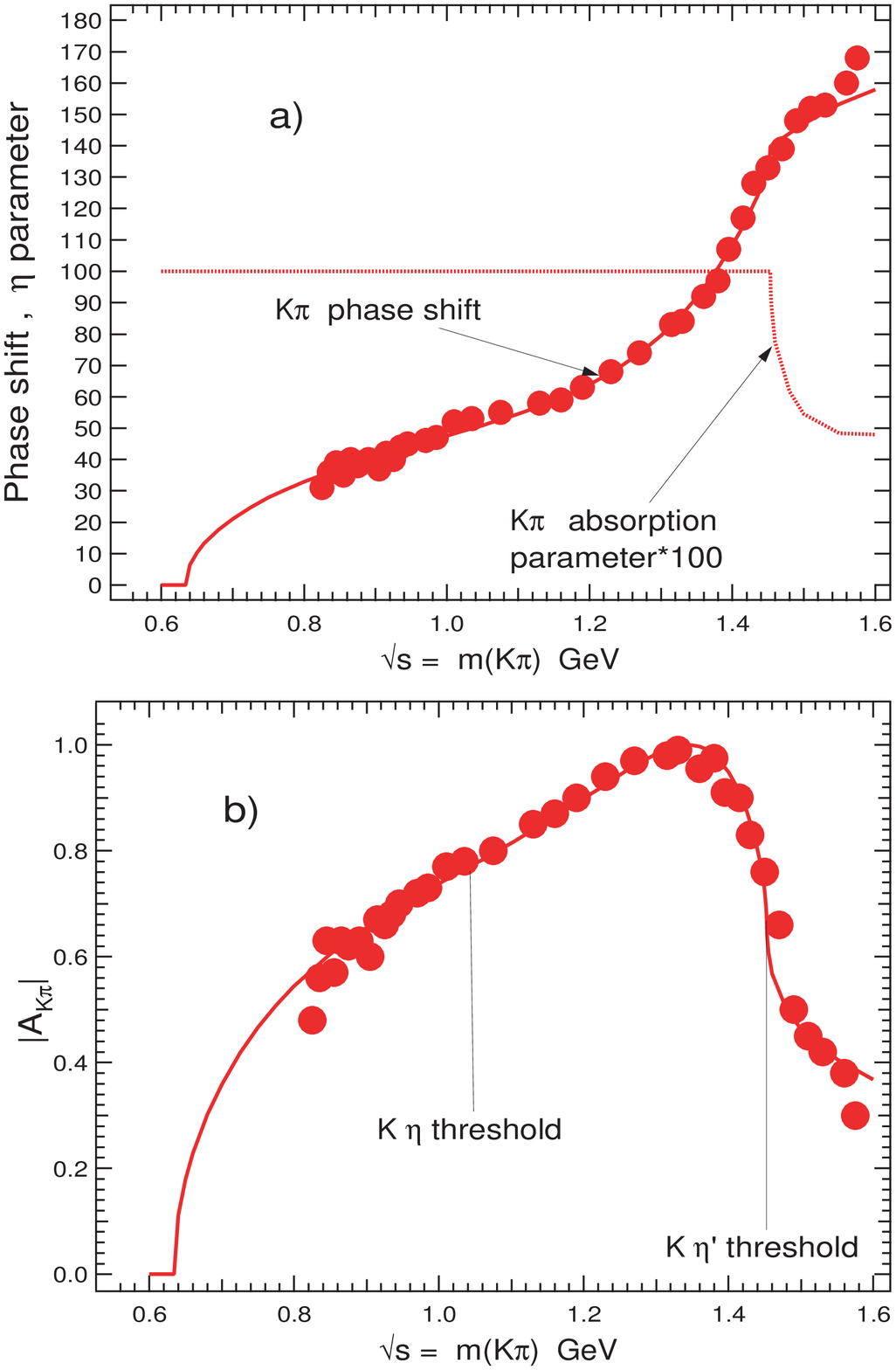}
\end{center}
\caption{\it The $K\pi$ S-wave phase shift and (b) the magnitude
of the $K\pi$ PWA compared with the model predictions, which fix 4
($\gamma$, $m_0+m_s$, $k_0$ and $s_{A,K\pi}$) of the 6
parameters.}
\end{figure}

\begin{figure}
\vspace{1.5cm}
\begin{center}
\includegraphics{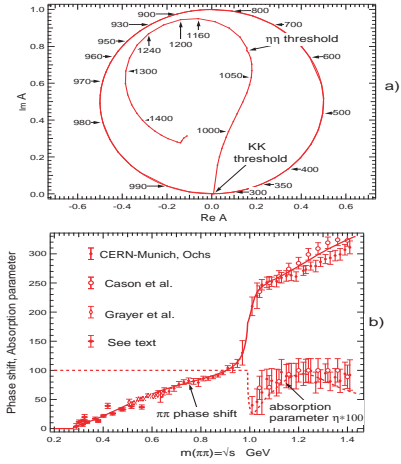}
\end{center}
\caption{\it  (a) The $\pi\pi$ Argand diagram and (b) phase shift
predictions are compared with data. Note that most of the
parameters were fixed by the data in Fig. 1. For more details see
Ref.~$^{1,2}$. }
\end{figure}

\begin{figure}
\vspace{3.4cm}
\begin{center}
\includegraphics{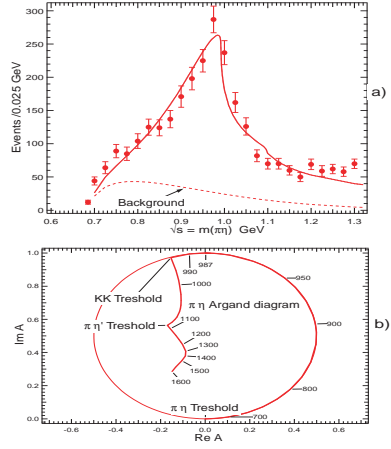}
\end{center}
\caption{\it (a) The $a_0(980)$ peak compared with model
prediction and (b) the predicted $\pi\eta$ Argand diagram.}
\end{figure}

\begin{figure}
\vspace{6cm}
\begin{center}
\includegraphics{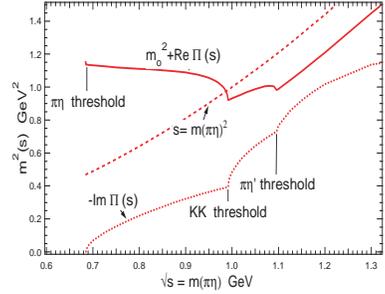}
\end{center}
\caption{\it The running mass $m_0+ Re\Pi(s)$ and $Im \Pi (s)$ of
the $a_0(980)$. The strongly dropping running mass at the
$a_0(980)$ position, below the $K\bar K$ threshold contributes to
the narrow shape of the peak in Fig. 3a.}
\end{figure}

\begin{figure}
\vspace{1cm}
\begin{center}
\includegraphics{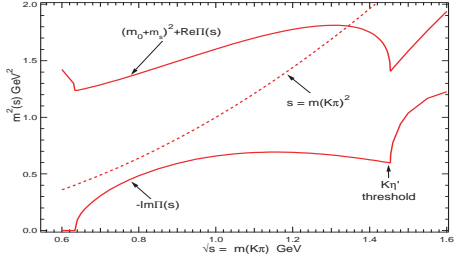}
\end{center}
\caption{\it The running mass and width-like function $-Im\Pi(s)$
for the $K^*_0(1430)$. The crossing of $s$ with the running mass
gives the 90$^\circ$ phase shift mass, which roughly corresponds
to a naive Breit-Wigner mass, where the running mass is put
constant.}
\end{figure}

\begin{figure}
\vspace{6cm}
\begin{center}
\includegraphics{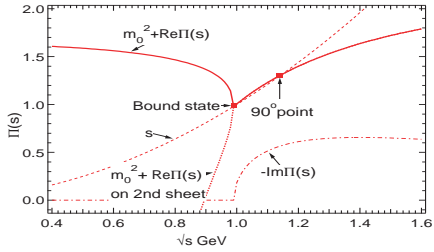}
\end{center}
\caption{\it (a) Although the model has only one bare $s\bar s $
resonance, when unitarized it can give rise to two crossings with
the running mass in the $s\bar s - K\bar K$ channels. This means
the $s\bar s$ state can manifest itself in two physical
resonances, one at threshold and one near 1200 MeV (See Ref.~$^2$
for details) as in this figure. }
\end{figure}

\begin{figure}
\vspace{0.5cm}
\begin{center}
\includegraphics{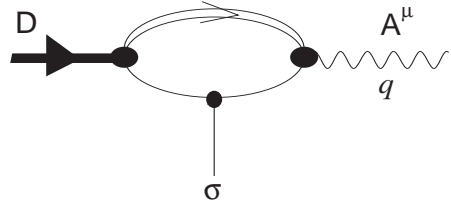}
\end{center}
\caption{\it Diagram for the {\it direct} contribution to the
$D\to\sigma$ semileptonic  amplitude. The axial current is
directly attached to the quark loop. }
\end{figure}

\begin{figure}
\vspace{3cm}
\begin{center}
\includegraphics{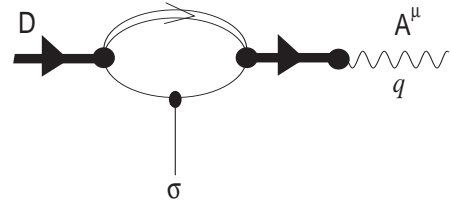}
\end{center}
\caption{\it The {\it polar} contribution to $F_0$, if a $0^-$
intermediate state is considered, and to $F_1$, with a $1^+$
intermediate state. The $D(1^+)$ state is described in the PDG
\cite{pdg2000}.}
\end{figure}

\begin{figure}
\vspace{3.6cm}
\begin{center}
\includegraphics{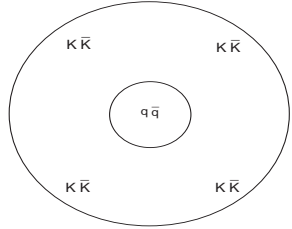}
\end{center}
\caption{\it A resonance just below the $K \bar K$ threshold,
coupling strongly to $K\bar K$, must have a large $K\bar K$
standing wave surrounding a $q\bar q$ core. }
\end{figure}

}

\end{document}